\documentclass[%
 reprint,
 longbibliography,
 amsmath,amssymb,
 aps,
 pra,
]{revtex4-1}

\usepackage{dsfont}
\usepackage{braket}
\usepackage{color}
\usepackage{graphicx}
\usepackage{dcolumn}
\usepackage{bm}

\begin{document}

\title{Particle emission from open quantum systems}

\author{Kevin A. Fischer}\email{kevinf@stanford.edu}
\affiliation{E. L. Ginzton Laboratory, Stanford University, Stanford CA 94305, USA}
\author{Rahul Trivedi}
\affiliation{E. L. Ginzton Laboratory, Stanford University, Stanford CA 94305, USA}
\author{Daniil Lukin}
\affiliation{E. L. Ginzton Laboratory, Stanford University, Stanford CA 94305, USA}

\date{\today}

\begin{abstract}
In this work, we discuss connections between different theoretical physics communities and their works, all related to systems that act as sources of particles such as photons, phonons, or electrons. Our interest is to understand how a low-dimensional quantum system driven by coherent fields, e.g. a two-level system, Jaynes--Cummings system, or photon pair source driven by a laser pulse, emits photons into a waveguide. Of particular relevance to solid-state sources is that we provide a way to include dissipation into the formalism for temporal-mode quantum optics. We will discuss the connections between temporal-mode quantum optics, scattering matrices, quantum stochastic calculus, continuous matrix product states and operators, and very traditional quantum optical concepts such as the Mandel photon counting formula and the Lindblad form of the quantum-optical master equation. We close with an example of how our formalism relates to single-photon sources with dephasing.
\end{abstract}

\maketitle



\section{Introduction}

\begin{figure}[b]
  \includegraphics[scale=0.80]{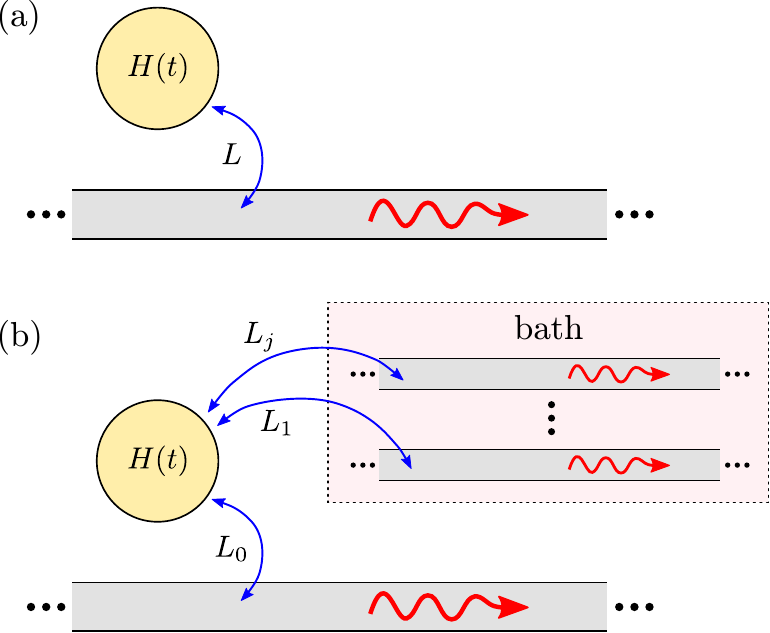}
  \caption{The general problem we solve in this manuscript is to compute the field scattered into a unidrectional (chiral) $1$-$d$ field (waveguide) from an energy-nonconserving $0$-$d$ Hamiltonian. This class of Hamiltonian is often used to represent coherent laser pulses scattering off quantum-optical systems such as a two-level system, Jaynes--Cummings system, or entangled photon pair source. First, we discuss just a single waveguide (a) and later extend to the same system coupled to a bath of modes that induce dissipation (b).}
  \label{fig:1}
\end{figure}

An open-quantum system consists of a local system, described by a low-dimensional ($0$-$d$) Hamiltonian $H$ acting on the Hilbert space $\mathbb{H}_\text{sys}$, coupled to one or more reservoirs or baths of modes in the Hilbert space $\mathbb{H}_\text{bath}$ via a spatially local coupling (Fig. \ref{fig:1}). In this work, we specifically consider the spatially local coupling in the Markovian or white-noise limit \cite{gardiner2004quantum}, which allows the baths to be decomposed into a set of quantum modes that have only a single continuous degree of freedom and hence are called $1$-$d$ fields (although the underlying mode function may have additional spatial dependence). The field operators are usually bosonic to represent photons or phonons, but occasionally emission into fermionic reservoirs is also considered. In the main text we will discuss the case with bosonic reservoirs, however, it is well known that the local system interacts bosonic and fermionic reservoirs identically. The same similarity is also true for the new formalism presented in this work (see Appendix A).

In these approximations, the open-quantum system may be fully described by just the Hamiltonian of the local system $H$ and a tuple of operator-rate products $\mathbf{L}$ from the local system that couple to the reservoir fields \cite{combes2017slh}. For our specific case, we consider the local system Hamiltonian to be time-varying $H\rightarrow H(t)$ so that it injects energy into a reservoir of interest, which we shall henceforth refer to as the waveguide. Such a situation corresponds physically to modeling a semi-classical coherent field driving the local system and causing it to scatter photons into the waveguide \cite{fischer2017pulsed,fischer2017scattering,shi2015multiphoton,caneva2015quantum,pletyukhov2015quantum,chang2016deterministic}. This is extremely important for modeling sources of nonclassical light \cite{michler,senellart2017high,fischer2017signatures,pavesi2016silicon}, and it was recently understood that quantum-optical systems can be used as auxiliary systems to generate one-dimensional continuous matrix product states (CMPS) \cite{barrett2013simulating,eichler2015exploring,pichler2017universal,gough2014quantum}. Hence, we take the state of the local system plus waveguide $\ket{\Psi(t)}\in\mathbb{H}_\text{sys}\otimes\mathbb{H}_\text{wg}$ at time $t=0$ to be $\ket{\Psi(0)}=\ket{\psi}\otimes \ket{\bm{0}}\equiv\ket{\psi,\bm{0}}$, i.e. with the waveguide in its vacuum state.

\begin{figure*}[t]
\includegraphics[scale=0.80]{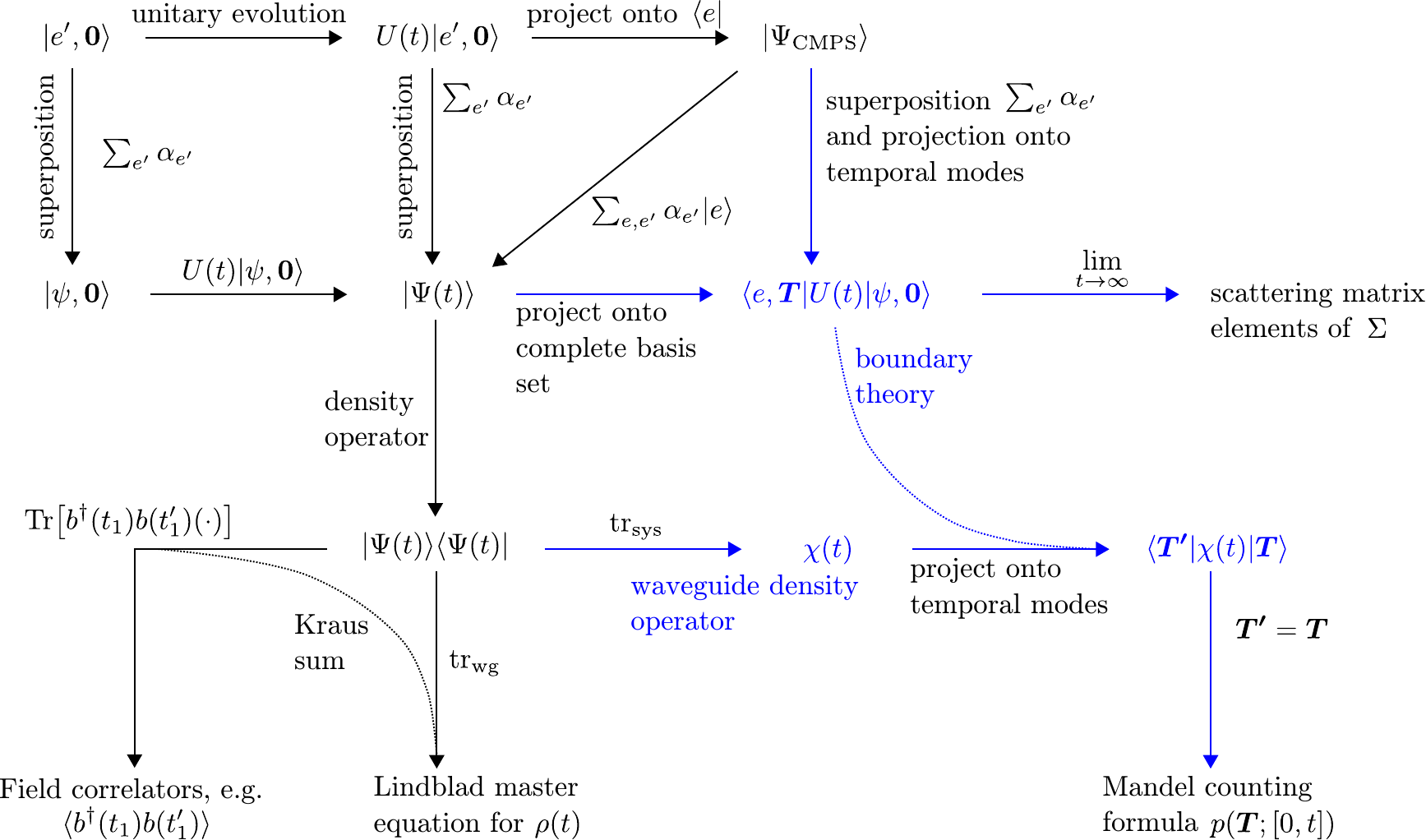}
  \caption{General relationship between concepts in the theory of particle emission into a waveguide. Starting from the initial state of $\ket{e',\bm{0}}$, which represents the system beginning in a local eigenstate $\ket{e'}$ and the waveguide in vacuum, the arrows indicate increased embedding or calculation based on the previous object. Although this diagram is drawn for the case of just a single waveguide, the relationships can easily be extended to include the additional loss channels of the bath (like in Fig. 1b) by changing $\text{tr}_\text{sys}\rightarrow\text{tr}_\text{sys}\text{tr}_\text{bath}$ and $\text{tr}_\text{wg}\rightarrow\text{tr}_\text{wg}\text{tr}_\text{bath}$ (the Kraus sum technique is also required). The dotted lines indicate that two concepts are connected via specific types of techniques discussed in the paper. Blue color indicates novel contributions of this work.}
  \label{fig:2}
\end{figure*}

Before beginning, we briefly outline the paper pictorially. For the system starting in an orthonormal basis state of the local system $\ket{e'}$ and the waveguide in vacuum ($\ket{\Psi(0)}=\ket{e',\bm{0}}$), Fig. 2 schematically shows how each of the objects and concepts we address are related to one another. Our main contributions related to arrows on this diagram are to:
\begin{itemize}
\item Relate CMPS to a boundary theory we develop from quantum stochastic calculus.
\item Show how the boundary theory can be used to obtain the scattering matrix of particle emission $\Sigma$.
\item Use the boundary theory to compute the waveguide density matrix by tracing out the system state, and in the presence of other loss channels called the bath (Fig. 1b) tracing them out as well.
\end{itemize}

\section{Emission into a single waveguide}

The total Hamiltonian can be written in terms of the local system Hamiltonian, the waveguide Hamiltonian, and their interaction: $H_\textrm{tot}=H(t)\otimes\mathds{1} + \mathds{1}\otimes H_{\textrm{wg}} + V$ (and is pictorially shown in Fig. \ref{fig:1}a). In an interaction picture with respect to the waveguide evolution
\begin{eqnarray}
\tilde{H}_\textrm{tot}(t)&=&\textrm{e}^{\textrm{\scriptsize i}H_{\textrm{wg}} t}H_\textrm{tot}\textrm{e}^{-\textrm{\scriptsize i}H_{\textrm{wg}} t}\nonumber\\
&=&H(t)\otimes\mathds{1}+\textrm{i}\left(L \otimes b^\dagger(t) - L^\dagger\otimes b(t)\right),\label{eq:Vt}
\end{eqnarray}
where $L$ is the product of a system operator $\sigma$ and a rate $\sqrt{\gamma}$. The operator $b(t)$ is the temporal mode operator for the waveguide, which obeys $[b(t),b^\dagger(s)]=\delta(t-s)$ with $b(t)\ket{\bm{0}}=0$, and hence creates a delta-normalized excitation of the waveguide in time. In terms of the frequency-mode annihilation operator $b(\omega)$ that annihilates a photon at a particular frequency, $b(t)=\int\mathop{\text{d}\omega}b(\omega)\text{e}^{-\text{i}\omega t}/\sqrt{2\pi}$. 

For a short time increment $\text{d}t$, the evolution operator of the interaction-picture wavefunction can be expanded in a Born approximation. Keeping terms to only $\mathcal{O}(\text{d}t)$ in the system and $\mathcal{O}(\sqrt{\text{d}t})$ in the waveguide gives rise to a quantum stochastic differential equation (QSDE)---these QSDE's allow for overcoming the singularities in the Schr\"{o}dinger equation from the temporal mode operators \cite{wiseman2009quantum,jacobs2014quantum,gardiner2015quantum,gardiner2004quantum,pan2016exact,dong2018controlling}. Specifically, the It\={o} increment for the unitary propagator that describes evolution of $\ket{\Psi(t)}$ over the interval $[t+dt)$ is given by (with $\hbar=1$)
\begin{subequations}
\begin{eqnarray}
\mathop{\text{d}U(t)}&=&U(t+\mathop{\text{d}t})-U(t)\\
&=&\Big\{-\text{i}H_\text{eff}(t)\otimes\mathds{1}\mathop{\text{d}t}\nonumber\\
&&\quad+L\otimes \mathop{\text{d}B^\dagger(t)}-L^\dagger\otimes\mathop{\text{d}B(t)}\Big\}U(t),\label{eq:qsde}
\end{eqnarray}
\end{subequations}
and hence the evolution can always be decomposed as
\begin{subequations}
\begin{eqnarray}
U(t)&\equiv& U(t,0)\\
&=&U(t,s_n)\cdots U(s_2,s_1)U(s_1,0)
\end{eqnarray}
\end{subequations}
for any $t>s_n>\dots>s_1>0$. The non-Hermitian effective `Hamiltonian' is $H_\text{eff}(t)=H(t)-\text{i}\frac{1}{2}L^\dagger L$ and the time-integrated quantity $B(t)=\int_0^{t} \mathop{\text{d}s} b(s)$ with $\text{d}B(t)=B(t+\mathop{\text{d}t})-B(t)$ is called the quantum noise or field increment. Hence,
\begin{equation}
\lim_{\text{d}t\rightarrow 0}\frac{\text{d}B(t)}{\text{d}t} = b(t),\label{eq:limdBt}
\end{equation}
the increments commute with each other for non-equal times, and $\text{d}B(t)\ket{\bm{0}}=0$. In the derivation of Eq. \ref{eq:qsde} and later to evaluate products of operators (involving $U(t)$ and) acting on vacuum, the zero-temperature It\={o} algebra is used:
\begin{center}
\begin{tabular}{ c|cccc } 
$\times$ & $\text{d}B(t)$ & $\text{d}B^\dagger(t)$ & $\text{d}t$ \\
\hline
$\text{d}B(t)$ & 0 & $\mathop{\text{d}t}$ & 0 \\ 
$\text{d}B^\dagger(t)$ & 0 & 0 & 0\\ 
$\text{d}t$ & 0 & 0 & 0\\ 
\end{tabular}
\end{center}

The formal solution to the evolution operator is given by integrating Eq. \ref{eq:qsde}
\begin{equation}
U(t) =  \mathcal{T}\text{e}^{\int_{0}^{t}\{-\text{i}H_\text{eff}(s)\otimes\mathds{1}\mathop{\text{d}s}+L \otimes \mathop{\text{d}B^\dagger(s)}-L^\dagger\otimes\mathop{\text{d}B(s)}\}},\label{eq:formalsoln}
\end{equation}
where $\mathcal{T}$ is the chronological operator that time-orders the infinitesimal products of Eq. \ref{eq:formalsoln}. Then, the wavefunction of the total waveguide and system at time $t$ is given by
\begin{eqnarray}
\ket{\Psi(t)}&=&U(t)\ket{\psi,\bm{0}}\nonumber\\
&=&\left(\mathds{1}\otimes\mathds{1}\right)U(t)\ket{\psi,\bm{0}}\nonumber\\
&=&\sum_e\int\mathop{\text{d}\bm{T}}\ket{e,\bm{T}}\bra{e,\bm{T}}U(t)\ket{\psi,\bm{0}}\label{eq:identity}
\end{eqnarray}
where $\ket{e,\bm{T}}\equiv\ket{e}\otimes\ket{\bm{T}}$ given $\{\ket{e}\}$ and $\{\ket{\bm{T}}\}$ form orthonormal bases for states in $\mathbb{H}_\text{sys}$ and $\mathbb{H}_\text{wg}$, respectively. To be concrete about the waveguide states, $\bm{T}=\{t_1,\dots,t_n\}$ is a time-ordered $N[\bm{T}]=n$ element vector that parameterizes the state $\ket{\bm{T}}=b^\dagger(t_1)\cdots b^\dagger(t_n)\ket{\bm{0}}$ and $\braket{\bm{T'}|\bm{T}}=\delta(\bm{T'}-\bm{T})$ \cite{fischer2017scattering,kiukas2015equivalence}. Hence, $\int\mathop{\text{d}\bm{T}}\equiv\sum_{n=0}^\infty\int_{0<t_1<\dots<t_n<t}\mathop{\text{d}t_1}\cdots\mathop{\text{d}t_n}$.
\vspace{5ex}

\subsection{Connection to scattering theory and CMPS}

We briefly relate this expansion to two important formalisms. First, when the Hamiltonian is asymptotically time independent and has at least one well-defined ground state, then the propagator becomes the scattering matrix $\Sigma$ \cite{fischer2017scattering} and its expansion in terms of the temporal modes is given by
\begin{eqnarray}
\bra{e,\bm{T}}\Sigma\ket{\psi,\bm{0}}=\lim_{t\rightarrow\infty}\bra{e,\bm{T}}U(t)\ket{\psi,\bm{0}}.
\end{eqnarray}
Let $t_c$ be the time when the Hamiltonian conserves energy again, then these scattering elements may be nonzero if $\lim_{t\rightarrow\infty} \mathcal{T}\text{e}^{-\text{i}\int_{t_c}^{t}\mathop{\text{d}s}H_\text{eff}(s)}\ket{e}$ has finite norm \cite{rahul}. Second, when $U(t)$ operates on vacuum, the result can be simplified to
\begin{subequations}
\begin{eqnarray}
\ket{\Psi(t)} &=&U(t)\ket{\bm{0}}\nonumber\\
&=&\mathcal{T}\text{e}^{\int_{0}^{t}\{-\text{i}H_\text{eff}(s)\otimes\mathds{1}\mathop{\text{d}s}+L \otimes \mathop{\text{d}B^\dagger(s)}\}}\ket{\bm{0}} \label{eq:UdB}\\
&=&\mathcal{T}\text{e}^{\int_{0}^{t}\mathop{\text{d}s}\{-\text{i}H_\text{eff}(s)\otimes\mathds{1}+L \otimes b^\dagger(s)\}}\ket{\bm{0}}.\label{eq:Ub}
\end{eqnarray}
\end{subequations}
This is done by first making use of the fact that $[\mathds{1}\otimes\text{d}B(t),U(t)]=0$ with $\text{d}B(t)\ket{\bm{0}}=0$ to remove the field annihilation operators \cite{gardiner2004quantum}. The equivalence of Eqs. \ref{eq:UdB} and \ref{eq:Ub} can be seen by expanding each exponential operator with a Dyson series and making use of the definition for $B(t)$. If $R=\ket{e_i}\bra{e_j}$, then a one-dimensional continuous matrix product state \cite{haegeman2013calculus,verstraete2010continuous,ganahl2017continuous} can be constructed from
\begin{eqnarray}
\ket{\Psi_\text{CMPS}}&=&\text{tr}_\text{sys}\left[R\,U(t)\right]\ket{\bm{0}} \\
&=&  \text{tr}_\text{sys}\left[R \,\mathcal{T}\text{e}^{\int_{0}^{t}\mathop{\text{d}s}\{-\text{i}H_\text{eff}(s)\otimes\mathds{1}+L \otimes b^\dagger(s)\}}\right]\ket{\bm{0}}.\nonumber
\end{eqnarray}
We note this is not the most general 1-$d$ CMPS---we would need to allow the coupling operator to vary in time $L\rightarrow L(t)$. These states were originally constructed by putting the continuum of modes from a Hamiltonian like Eq.~\ref{eq:Vt} on a lattice, and then taking the thermodynamic limit where the lattice spacing vanishes \cite{osborne2010holographic}. This turns out to be mathematically equivalent to the coarse-graining-in-time action of the quantum stochastic calculus we use here.

\subsection{Boundary theory}

Evaluating the propagator has previously been reduced to calculating expectations of system operators, through various different means (e.g. \cite{haegeman2013calculus,jennings2015continuum,fischer2017scattering,shi2015multiphoton,rahul,xu2017input,caneva2015quantum,pan2016exact}. In field theory, this result is referred to as the holographic property of CMPS \cite{osborne2010holographic} and in quantum optics language, we call this a result of the boundary condition from input-output theory \cite{gardiner2004quantum}. These formulations have been covered extensively, and we will arrive at something like the CMPS or scattering matrix result but using the language of quantum stochastic differential equations.

Our next step in reducing the complexity of this problem is to turn the expansion of $U(t)$ into vacuum expectation values \cite{fischer2017scattering}. To do this, we need the commutation $[\mathds{1}\otimes\mathop{\text{d}B(s)},\text{d}U(s)]=\left(L\otimes\mathds{1}\right)\mathop{\text{d}t}U(s)$ and the limit from Eq.~\ref{eq:limdBt}, which together give us the relation
\begin{widetext}
\begin{eqnarray}
[\mathds{1}\otimes b(t_1),U(s,0)]&=&\lim_{\text{d}t\rightarrow 0}[\mathds{1}\otimes \mathop{\text{d}B(t_1)}/\mathop{\text{d}t},U(s,t_1+\mathop{\text{d}t})\left(\mathop{\text{d}U(t_1)}+U(t_1, 0)\right)]\nonumber\\
&=&\lim_{\text{d}t\rightarrow 0}U(s,t_1+\mathop{\text{d}t})\left[\mathds{1}\otimes\mathop{\text{d}B(t_1)},\mathop{\text{d}U(t_1)} \right]/\mathop{\text{d}t}\nonumber\\
&=&U(s,t_1)\left(L\otimes\mathds{1}\right)U(t_1,0)\label{eq:commutestuff}
\end{eqnarray}
\end{widetext}
if $s>t_1>0$. Using this commutation and the fact that $b(t)\ket{\bm{0}}=0$, we remove the free field annihilation operators from the expectation
\begin{eqnarray}
\bra{e,\bm{T}}U(t)\ket{\psi,\bm{0}}=&\nonumber\\
&\hspace{-14ex}\bra{e,\bm{0}}U(t,t_n)\left(L\otimes\mathds{1}\right)U(t_{n},t_{n-1})\left(L\otimes\mathds{1}\right)\cdots \nonumber\\&\left(L\otimes\mathds{1}\right)U(t_1,0)\ket{\psi,\bm{0}}.
\end{eqnarray}
We note this expression is written in the temporally factorized form 
\begin{eqnarray}
\bra{\bm{0}}A(t,s+\mathop{\text{d}t})U(s+\mathop{\text{d}t},s)C(s,0)\ket{\bm{0}}
\end{eqnarray}
for arbitrary $s$. The field increments from $U(s+\mathop{\text{d}t},s)$ commute towards the vacuum states and annihilate, given that $[\text{d}B(s),C(s,0)]=0$ and $[\text{d}B^\dagger(s),A(t,s+\mathop{\text{d}t})]=0$. Hence, in the vacuum expectation the unitary evolution operators cannot create or annihilate particles and we make the replacement $\mathop{\text{d}U(s)}\rightarrow \{-\text{i}H_\text{eff}(s)\mathop{\text{d}s}\otimes\mathds{1}\}U(s)$ for all $s$. Then we define
\begin{eqnarray}
V(t_1,t_0) &=&\bra{\bm{0}}U(t_1,t_0)\ket{\bm{0}}\nonumber\\
&=&\mathcal{T}\text{exp}\left[-\text{i}\int_{t_0}^{t_1}\mathop{\text{d}s}H_\text{eff}(s)\right]
\end{eqnarray}
and write the expectation value in terms of only system operators
\begin{eqnarray}
\bra{e,\bm{T}}U(t)\ket{\psi,\bm{0}}=&\label{eq:boundary}\\
&\hspace{-15ex}\bra{e}V(t,t_n)LV(t_{n},t_{n-1})L\cdots LV(t_1,0)\ket{\psi}.\nonumber
\end{eqnarray}
This expectation has a very intuitive form, where the non-unitary propagators $V(\cdot)$ correspond to evolution conditioned on no particle emission into the field, and the $L$ operators scatter a particle into the waveguide. (This result is similar as we derived in Refs. \cite{fischer2017scattering,rahul}---there we also noted that the pure-state calculation of Eq. \ref{eq:boundary} need only be performed until the time $t_\text{c}$, when energy is again conserved, and projected onto the local system's ground states.)

We can also write the waveguide state's $U(\cdot)$ evolution as a density matrix $\chi(t)=\text{tr}_\text{sys}\left[\ket{\Psi(t)}\bra{\Psi(t)}\right]$, with $\chi(0)=\ket{\bm{0}}\bra{\bm{0}}$. This calculation is a bit unusual: rather than tracing over the waveguide, it is much more standard to trace out the system and obtain a quantum-optical master equation (Sec. \ref{sec:me}). Expanding this density matrix in the temporal mode basis $\braket{\bm{T'}|\chi(t)|\bm{T}}=\text{Tr}\left[\ket{\bm{T}}\braket{\bm{T'}|\Psi(t)}\bra{\Psi(t)}\right]$ and utilizing Eq. \ref{eq:identity} and Eq. \ref{eq:boundary} twice (once for the bra $\bra{\Psi(t)}=\bra{\Psi(0)}U^\dagger(t)$ and once for the ket $\ket{\Psi(t)}=U(t)\ket{\Psi(0)}$), yields
\begin{eqnarray}
\braket{\bm{T'}|\chi(t)|\bm{T}} =& \label{eq:chi_t1}\\
&\hspace{-5ex}\text{tr}_\text{sys}\big[\mathcal{V}(t,\tilde{\tau}_R)\mathcal{S}_{Q[\tilde{\tau}_{R}]}\mathcal{V}(\tilde{\tau}_{R},\tilde{\tau}_{R-1})\mathcal{S}_{Q[\tilde{\tau}_{R-1}]}\nonumber\\
&\hspace{14ex}\cdots\mathcal{S}_{Q[\tilde{\tau}_{1}]}\mathcal{V}(\tilde{\tau}_1,0)\ket{\psi}\bra{\psi}\big],\nonumber
\end{eqnarray}
where we define a chronologically sorted list of times $\{\tilde{\tau}_1,\dots,\tilde{\tau}_R\}=\text{sort}\{\bm{T'}+\bm{T}\}$ and $Q[\tilde{\tau}]\in\{0,1\}$ depending on whether the time came from $\bm{T'}$ or $\bm{T}$. We also use a script letter to mean a superoperator, where $\mathcal{V}(t,0)\chi\equiv V(t,0) \chi V(0,t)$, $\mathcal{S}_0\chi=L\chi$, and $\mathcal{S}_1\chi=\chi L^\dagger$. As a reminder, Eq. \ref{eq:chi_t1} describes the field state but is written in terms of \textit{system} superoperators only. Such a density matrix has been coined both a matrix product operator \cite{pirvu2010matrix,kiukas2015equivalence} or superoperator state \cite{grimsmo2015time}.

\subsection{Master equation\label{sec:me}}

On the other hand, the quantum-optical master equation for the reduced dynamics of the system is obtained by applying unitary evolution and tracing out the waveguide degrees of freedom \cite{wiseman2009quantum,gardiner2004quantum,jacobs2014quantum}, often using the Kraus sum operator technique
\begin{subequations}\label{eq:rho_t}
\begin{eqnarray}
\rho(t) &=& \text{tr}_\text{wg}\big[\ket{\Psi(t)}\bra{\Psi(t)}\big]\\
&=&\text{tr}_\text{wg}\big[\mathcal{U}(t,0)\{\ket{\psi}\bra{\psi}\otimes\ket{\bm{0}}\bra{\bm{0}}\}\big].
\end{eqnarray}
\end{subequations}
Here, $\mathcal{U}(t,0)\rho\equiv U(t,0) \rho\, U(0,t)$ is the unitary evolution superoperator. Further, because the system-waveguide coupling is Markovian Eq. \ref{eq:rho_t} can always be written as
\begin{eqnarray}
\rho(t_1)&=&\text{tr}_\text{wg}\big[\mathcal{U}(t_1,t_0)\{\rho(t_0)\otimes\ket{\bm{0}}\bra{\bm{0}}\}\big],
\end{eqnarray}
or similarly in the form of a Liouville equation $\dot{\rho}(t)=\mathcal{L}(t)\rho(t)$. Here, the Liovillian $\mathcal{L}(\cdot)$ superoperator (or transfer matrix $\mathbb{T}(\cdot)$ in CMPS papers) is defined by 
\begin{subequations}
\begin{eqnarray}
\mathcal{L}(t)\rho&=&-\text{i}[H(t), \rho] +\big\{-\tfrac{1}{2}L^\dagger L,\rho\big\} +L\rho L^\dagger\\
&=&-\text{i}[H_\text{eff}(t), \rho] +\mathcal{J}[L]\rho
\end{eqnarray}
\end{subequations}
where $\mathcal{J}[L]\rho=L\rho L^\dagger=\mathcal{S}_0\mathcal{S}_1\rho$ is the recycling or emission superoperator. We formally will express such a time evolution in terms of the superoperator $\mathcal{M}(\cdot)$ as
\begin{subequations}
\begin{eqnarray}
\rho(t_1)&=&\mathcal{M}(t_1,t_0)\rho(t_0)\\
&=&\mathcal{T} \text{exp}\left[\int_{t_0}^{t_1} \mathop{\text{d}s}\mathcal{L}(s)\right]\rho(t_0).
\end{eqnarray}
\end{subequations}

\section{Addition of loss channels\label{sec:loss}}

Our main contribution in this work is to formalize the effects of loss into other channels, and on how it causes the waveguide to enter a mixed state (shown pictorially in Fig. \ref{fig:1}b). Suppose $L_0$ represents coupling to the waveguide, whose state we want to keep track of, and the operators $L_1,\cdots,L_j$ represent coupling to other loss channels or the bath we will trace over. Then,
\begin{eqnarray}
\mathop{\text{d}U(t)}=\Big\{&\hspace{-23ex}-\text{i}H_\text{eff}(t)\otimes\mathds{1}\mathop{\text{d}t}\\
&+\sum_k L_k\otimes \mathop{\text{d}B_k^\dagger(t)}-L_k^\dagger\otimes \mathop{\text{d}B_k(t)}\Big\}U(t)\nonumber
\end{eqnarray}
where the field increments from separate channels trivially commute and now
\begin{eqnarray}
H_\text{eff}(t)=H(t)-\text{i}\sum_k\frac{1}{2}L_k^\dagger L_k.
\end{eqnarray}

\subsection{Waveguide field density operator}

Here, we need to use a density operator to keep track of the waveguide state
\begin{eqnarray}
\chi(t) &=& \text{tr}_\text{sys}\text{tr}_\text{bath}\big[\ket{\Psi(t)}\bra{\Psi(t)}\big]\\
&=&\text{tr}_\text{sys}\text{tr}_\text{bath}\big[\mathcal{U}(t,0)\{\ket{\psi}\bra{\psi}\otimes\ket{\bm{0}}\bra{\bm{0}}_\text{wg}\ket{\bm{0}}\bra{\bm{0}}_\text{bath}\}\big].\nonumber
\end{eqnarray}
Again, projecting $\chi(t)$ onto the temporal mode basis like in Eq. \ref{eq:chi_t1}
\begin{widetext}
\begin{eqnarray}
\braket{\bm{T'}|\chi(t)|\bm{T}} = \label{eq:chi_projected}\\
\text{tr}_\text{sys}\text{tr}_\text{bath}\big[&\mathcal{V}(t,\tilde{\tau}_R)\left(\mathcal{S}_{Q[\tilde{\tau}_{R}]}\otimes\mathds{1}_\text{bath}\right)\mathcal{V}(\tilde{\tau}_{R},\tilde{\tau}_{R-1})\left(\mathcal{S}_{Q[\tilde{\tau}_{R-1}]}\otimes\mathds{1}_\text{bath}\right)\cdots\left(\mathcal{S}_{Q[\tilde{\tau}_{1}]}\otimes\mathds{1}_\text{bath}\right)\mathcal{V}(\tilde{\tau}_1,0)\{\ket{\psi}\bra{\psi}\otimes\ket{\bm{0}}\bra{\bm{0}}_\text{bath}\}\big]\nonumber
\end{eqnarray}
\end{widetext}
and noting $\mathcal{S}_0\chi=L_0\chi$ and $\mathcal{S}_1\chi=\chi L^\dagger_0$,
but now with $V(t_1,t_0) =\bra{\bm{0}}U(t_1,t_0)\ket{\bm{0}}_\text{wg}$ or
\begin{eqnarray}
V(t_1,t_0) &=&\mathcal{T}\text{exp}\Bigg[\displaystyle\int_{t_0}^{t_1}\Big\{-\text{i}H_\text{eff}(s)\otimes\mathds{1}_\text{bath}\mathop{\text{d}s}\qquad\\
&&\hspace{14ex}+\sum_{k>0} L_k\otimes \mathop{\text{d}B_k^\dagger(t)}\Big\}\Bigg].\nonumber
\end{eqnarray}
Taking the trace over the bath state, which can easily be done according to the standard rules of quantum stochastic calculus,
\begin{eqnarray}
\braket{\bm{T'}|\chi(t)|\bm{T}} = &\label{eq:chi_tpt}\\
&\hspace{-10ex}\text{tr}_\text{sys}\big[\mathcal{K}(t,\tilde{\tau}_R)\mathcal{S}_{Q[\tilde{\tau}_{R}]}\mathcal{K}(\tilde{\tau}_{R},\tilde{\tau}_{R-1})\mathcal{S}_{Q[\tilde{\tau}_{R-1}]}\nonumber\\
&\hspace{15ex}\cdots\mathcal{S}_{Q[\tilde{\tau}_{1}]}\mathcal{K}(\tilde{\tau}_1,0)\ket{\psi}\bra{\psi}\big].\nonumber
\end{eqnarray}
Here,
\begin{eqnarray}
\mathcal{K}(t_1,t_0)&=&\text{tr}_\text{bath}\big[\mathcal{V}(t_1,t_0)\{\ket{\psi}\bra{\psi}\otimes\ket{\bm{0}}\bra{\bm{0}}_\text{bath}\}\big]\nonumber\\
&=&\mathcal{T} \text{exp}\left[\int_{t_0}^{t_1} \mathop{\text{d}s}\big\{\mathcal{L}(s)-\mathcal{J}[L_0]\big\}\right]
\end{eqnarray}
which can be thought of as an unnormalized map that evolves the density matrix conditional on no photon emissions into the $0$-th reservoir and
\begin{eqnarray}
\mathcal{L}(t)\rho&=&-\text{i}[H(t), \rho] +\sum_k\big\{-\tfrac{1}{2}L_k^\dagger L_k,\rho\big\} +L_k\rho L_k^\dagger\nonumber\\
&=&-\text{i}[H_\text{eff}(t), \rho] +\sum_k\mathcal{J}[L_k]\rho
\end{eqnarray}
is the new Liouvillian including all $L_0,L_1,\dots,L_j$. This new $\mathcal{L}(\cdot)$ with inclusion of the bath is now the generator of the map $\mathcal{M}(\cdot)$. It is fairly trivial to extend this work to cases where the bath is in a thermal state, by using a different set of It\={o} algebra \cite{gardiner2004quantum}---we simply opted for a more economical exposition here. We now have access to the entire state of the waveguide: we will later calculate quantities such as the trace purity of the emitted states, which is of interest for few-photon sources.

\subsection{Particle counting formula}

If $\bm{T'}=\bm{T}$, then $\braket{\bm{T}|\chi(t)|\bm{T}}$ gives precisely the Mandel counting formula \cite{gardiner2015quantum,carmichael2009open} for the probability density of $n$ particle emissions to occur at the times $t_1,\dots,t_n$ within the interval $[0,t]$, i.e.
\begin{eqnarray}
p(t_1,\dots,t_n;[0,t])=\braket{\bm{T}|\chi(t)|\bm{T}}
\end{eqnarray}
or equivalently
\begin{eqnarray}
p(\bm{T};[0,t])=\text{tr}_\text{sys}\big[\mathcal{K}(t,t_n)\mathcal{J}[L_0]\mathcal{K}(t_n,t_{n-1})\mathcal{J}[L_0]\cdots\nonumber\\
\mathcal{J}[L_0]\mathcal{K}(t_1,0)\ket{\psi}\bra{\psi}\big].\nonumber\\\label{eq:pn_l0}
\end{eqnarray}
The photocount distribution, i.e. the probability that $n$ particles are emitted is given by
\begin{eqnarray}
P_n=\int_{N[\bm{T}]=n}\mathop{\text{d}\bm{T}}p(\bm{T};[0,t]).\label{eq:Pn__0}
\end{eqnarray}

\subsection{Correlations between field operators\label{sec:corre_field}}

\begin{figure*}
  \centering
  \includegraphics[scale=0.90]{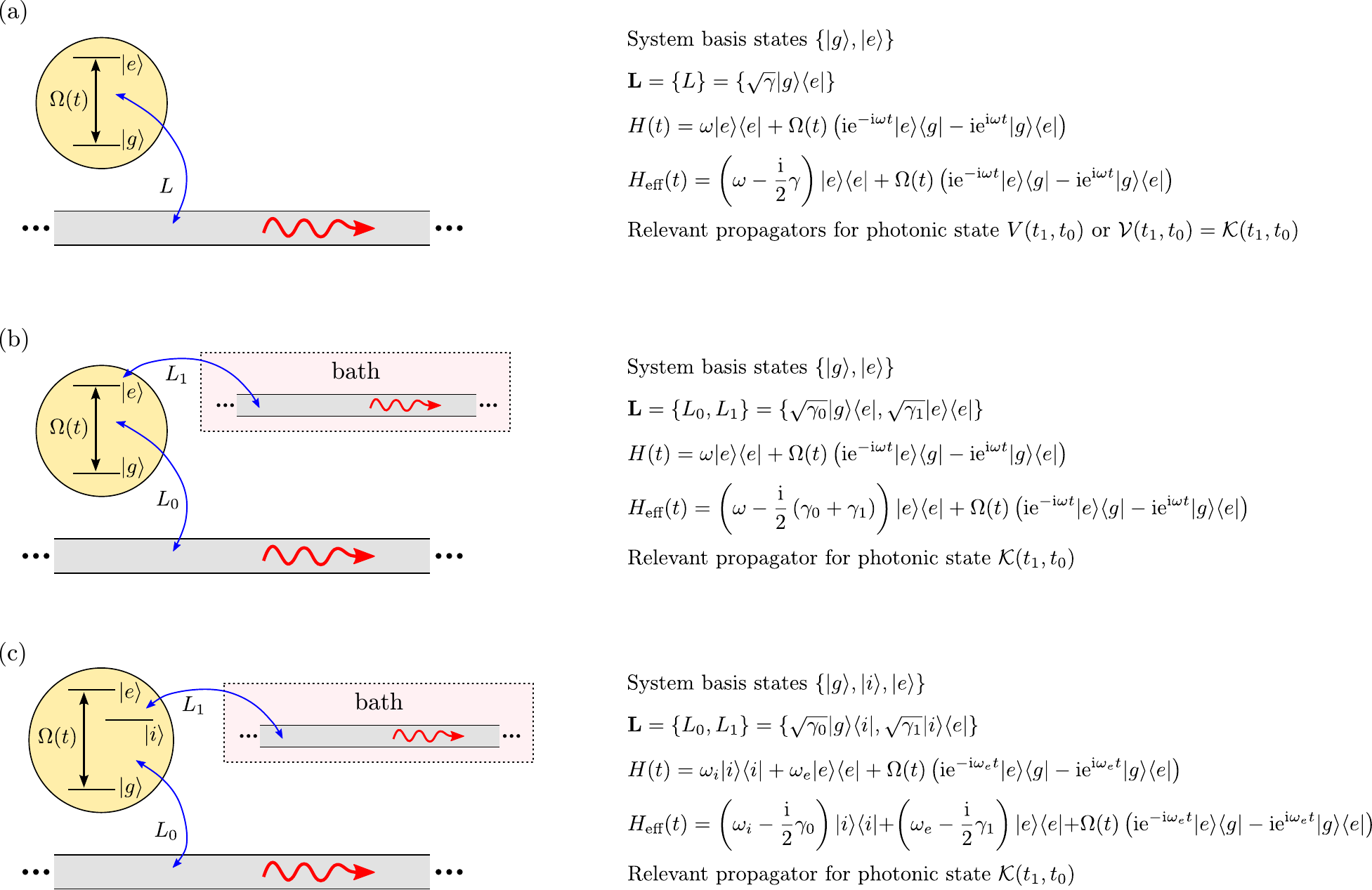}
  \caption{Examples of local quantum systems as particle sources, with their $(\textbf{L},H)$ tuples, corresponding non-Hermitian effective Hamiltonians, and relevant propagators for photon emission. (a) Two-level system, driven by a laser pulse. Because there is no dissipation, the scattered photonic state can be written and solved either as a pure-state $\ket{\Psi}$ or as a density matrix $\chi$. (b) Two-level system dispersively coupled to a bath, causing dephasing, and driven by a laser pulse. (c) Three-level cascade, driven by a laser pulse, where only emission from the transition $\ket{i}\rightarrow\ket{g}$ is of interest. Due to the dissipation for (b) and (c), the photonic state can only be a density matrix $\chi$. All laser pulses are resonant with the $\ket{g}\leftrightarrow\ket{e}$ transitions in (a)-(c).}
  \label{fig:3}
\end{figure*}

While it is very useful to compute the precise field state for understanding how a system emits light into the waveguide, often the only measurable information about the state comes from its normally- and time-ordered correlation functions
\begin{equation}
\braket{b^\dag(t_1)\cdots b^\dag(t_n)b(t_{n'}')\cdots b(t_1')}.
\end{equation}
In quantum optics, these correlation functions are almost always computed by using the boundary condition from input-output theory to relate the field correlations to correlations between system operators \cite{gardiner2004quantum}. Here, we use our quantum stochastic techniques.

Consider the (first-order coherence or) field-field correlator
\begin{subequations}
\begin{eqnarray}
G^{(1)}(t_1,t_1')&\equiv&\braket{b^\dagger(t_1)b(t_1')}\label{eq:G1_1}\\
&=&\text{Tr}\big[b^\dagger(t_1)b(t_1')\ket{\Psi(t)}\bra{\Psi(t)}\big]\label{eq:G1_2}\\
&=&\text{Tr}\big[b(t_1')\ket{\Psi(t)}\bra{\Psi(t)}b^\dagger(t_1)\big].\label{eq:G1_cyc}
\end{eqnarray}
\end{subequations}
Equations \ref{eq:G1_1} and \ref{eq:G1_2} are simply definitions, while Eq. \ref{eq:G1_cyc} is arrived at via the cyclic property of the trace. For this correlation to be nonzero, $t>t_1,t_1'$. Considering the specific case where $t>t_1>t_1'$ 
\begin{widetext}
\begin{eqnarray}
G^{(1)}(t_1,t_1')&=&\text{Tr}\big[\mathcal{U}(t,t_1)\left( \mathcal{S}_1\otimes\mathds{1}\right)\mathcal{U}(t_1,t_1')\left( \mathcal{S}_0\otimes\mathds{1}\right)\mathcal{U}(t_1',0)\{\ket{\psi}\bra{\psi}\otimes\ket{\bm{0}}\bra{\bm{0}}\}\big]\nonumber\\
&=&\text{Tr}\big[\left( \mathcal{S}_1\otimes\mathds{1}\right)\mathcal{U}(t_1,t_1')\left( \mathcal{S}_0\otimes\mathds{1}\right)\mathcal{U}(t_1',0)\{\ket{\psi}\bra{\psi}\otimes\ket{\bm{0}}\bra{\bm{0}}\}\big]\nonumber\\
&=&\text{tr}_\text{sys}\big[\mathcal{S}_1\mathcal{M}(t_1,t_1')\mathcal{S}_0\mathcal{M}(t_1',0)\ket{\psi}\bra{\psi}\big]\label{eq:G1_corr}.
\end{eqnarray}
\end{widetext}
(For the case where $t_1<t_1'$, consider that $G^{(1)}(t_1,t_1')$ is conjugate symmetric with respect to exchanging times.) The first step makes use of exactly the same commutation techniques as in Eqs. \ref{eq:chi_t1} and \ref{eq:chi_projected}. The second step is made by noting that unitary evolution preserves the trace of the density matrix so we replace $\mathcal{U}(t,t_1)\rightarrow\mathds{1}\otimes\mathds{1}$. The final state is an example application of the so-called quantum regression theorem, where $\mathcal{M}(\cdot)$ is again a map from the generator $\mathcal{L}(\cdot)$ including bath dissipation.

From this expression, it is clear that states with one particle \textit{or more} all contribute to the first-order coherence. If the number of particles emitted is small, however, then $\mathcal{M}(\cdot)\approx\mathcal{K}(\cdot)$ and the first-order coherence roughly gives the density matrix for a single-particle state in the waveguide $G^{(1)}(t_1,t_1')\approx\braket{t_1'|\chi(t)|t_1}$. Higher-order coherences such as the (second-order coherence or) intensity-intensity correlator can similarly be expressed in terms of system operators (e.g. take $t>t_2>t_1$)
\begin{eqnarray}
G^{(2)}(t_1,t_2)&=&\braket{b^\dagger(t_1)b^\dagger(t_2)b(t_2)b(t_1)}\\
&=&\text{tr}_\text{sys}\big[\mathcal{J}[L_0]\mathcal{M}(t_2,t_1)\mathcal{J}[L_0]\mathcal{M}(t_1,0)\ket{\psi}\bra{\psi}\big]\nonumber.
\end{eqnarray}


\section{Examples of local quantum systems as particle sources\label{sec:examples}}

In the previous sections, we overviewed mathematical methods for describing the scattered particle field from a local quantum system. For this section, we provide explicit examples of local systems that emit particles, and in particular show how a few common single-photon sources map onto the mathematics described in this paper. The streams of isolated photons produced by these sources are of interest as inputs to optical networks for quantum information processing \cite{michler,senellart2017high,loredo2017boson,o2009photonic,pavesi2016silicon}.

As we mentioned in the introduction, the local systems we consider act as sources of particles because their Hamiltonians are time-dependent. In the case of single-photon sources, the time-dependence typically results from a classical laser field driving the system. Because this work only considers open-quantum systems with Markovian coupling to the bath, the laser field can be included into the local system's Hamiltonian via a Mollow transformation. Then, $H(t)=H_0+H_1(t)$ where $H_0$ is the local system's undriven Hamiltonian and $H_1(t)=\text{i}\left(L\alpha^*(t)-L^\dag\alpha(t)\right)$ with $\alpha(t)$ as the amplitude of the coherent state. The effective driving strength of the laser on the system is typically written as the product of a real-valued envelope centered around the carrier frequency of the pulse and the coupling $L$, with
\begin{equation}
H_1(t)=\Omega(t)\left(\text{i}\text{e}^{\text{i}\omega t}\sigma^\dag-\text{i}\text{e}^{-\text{i}\omega t}\sigma\right)
\end{equation}
where $\sigma$ again is a system coupling operator.

Figure \ref{fig:3} shows three such example systems, along with their mathematical representations and mapping to the relevant propagators for photon emission discussed in this work. The first is the most basic system that can act as a source of single photons: the quantum two-level system driven by a short optical pulse from a laser (a) \cite{fischer2017scattering}. Two-level systems based on InGaAs quantum dots are some of the most promising single-photon sources, demonstrating the lowest multi-photon error rates, highest efficiencies, fastest generation speeds, and highest photon indistinguishabilities to date \cite{senellart2017high}. However, the physical implementations further suffer from various sources of dephasing, due to interactions with phonon and electron reservoirs  \cite{kuhlmann2015transform,iles2017phonon,luker2017phonon,gawarecki2012dephasing,wei2014deterministic,michler,reiter2014role,quilter2015phonon}. The simplest model including dephasing to the two-level system is shown in (b). Despite the experimental success of the two-level system as a single-photon source, the multi-photon error rates scale poorly with the pulse length. Recent investigations have shown that a 3-level photon cascade (c), where the photon is collected from just one transition of the cascade, has significantly lower multi-photon error rate \cite{schweickert2017demand,Kai,fischer2016dynamical}.

Hence, we have established the connection for relatively simple local quantum systems that emit particles and our mathematics. Lastly, we note that our formalism could be used to compute the scattered photonic state emitted from a network of open-quantum systems. In particular, by assigning each element in the network a scattering matrix $\mathbf{S}$ that describes its connections to other components, a single operator triple $(\textbf{S},\textbf{L},H)$ can be computed that gives a relationship between the input and output channels of the network (SLH theory, explicitly named) \cite{combes2017slh,gough2009series,gough2009quantum,gough2010squeezing,zhang2011direct,hamerly2012advantages}. The final $\textbf{L}$ and $H$ can then be used to construct the appropriate propagators $V(\cdot)$, $\mathcal{V}(\cdot)$, or $\mathcal{K}(\cdot)$ for particle emission described in this work.


\section{Application to photon indistinguishability}

In Sec. \ref{sec:loss}, we presented a new formalism for understanding how a local quantum system emits particles in the presence of loss. We saw how our formalism allows for the calculation of the photonic density matrix $\chi(t)$ in a waveguide, having traced over the unmonitored loss channels causing dissipation. Now, we explore how having access to elements of the photonic density matrix directly provides new information compared with previous approaches of analyzing the indistinguishability of single-photon sources.

For most applications requiring a source of single photons, the output photonic state must be as pure as possible (in the sense of trace purity). When the single-photons are in identical quantum-mechanically pure states, then they are indistinguishable and it is possible for two-photon interference to occur in a Hong--Ou--Mandel experiment. From this experiment, one can extract a parameter called the visibility
\begin{eqnarray}
v = \frac{\iint\mathop{\text{d}t_1}\mathop{\text{d}t_1'}\left| G^{(1)} \left(t_1, t_1' \right) \right|^2}{\braket{n}^2} ,
\end{eqnarray}
where $\braket{n}=\sum_n n P_n=\int\mathop{\text{d}t_1} G^{(1)} \left(t_1, t_1 \right)$ is the expected number of photons in the emitted pulse and $G^{(1)} \left(t_1, t_1' \right)$ is defined in Eq. \ref{eq:G1_corr} \cite{kiraz2004quantum,pathak2010coherently,fischer2016dynamical}. This parameter is traditionally used as a way to access the trace purity of the photons emitted by a single-photon source, where for perfect single photons $v=1$ while completely mixed single photons yield $v=0$.

\begin{figure}
  \includegraphics[]{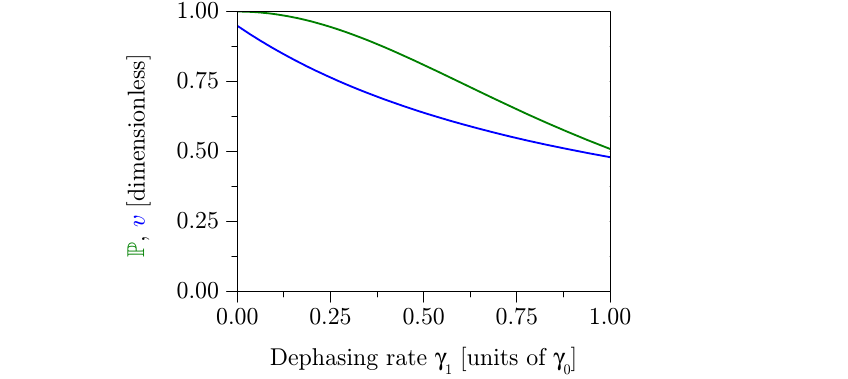}
  \caption{Photon emission from a two-level system with dephasing under excitation by a short laser pulse (shown schematically in Fig.~\ref{fig:3}b). The laser pulse has a temporal width $\tau_\text{p}=0.1/\gamma_0$ that injects approximately one quanta of energy into the photon field, $P_1=0.95$. Trace purity $\mathbb{P}$ of single-photon component of emission (solid green curve) and Hong--Ou--Mandel (HOM) interference parameter $v$ (solid blue curve) are shown.}
  \label{fig:4}
\end{figure}

By considering the quantum two-level system, coupled to a reservoir causing dephasing and driven by a short laser pulse (Fig. \ref{fig:3}b), we now show where this metric can fall short in estimating the trace purity. In particular, consider the excitation scenario of a square laser pulse where the length is one tenth of the spontaneous emission lifetime. This corresponds to a fairly high probability of single photon emission of $P_1=0.95$. For zero dephasing, the emitted state is entirely pure (in the trace sense) since the evolution to compute the final state vector is only the Schr\"odinger evolution, yet the visibility $v<1$ (Fig. \ref{fig:4} blue curve for $\gamma_1=0$). The reason that $v\neq1$ despite being a quantum-mechanically pure state is that the first-order coherence, and hence the HOM visibility, is sensitive to multiple photons not just a single one (as discussed in Sec. \ref{sec:corre_field}).

Because our formalism for understanding particle emission provides direct access to the density matrix of the photonic state $\chi(t)$, it is now possible to calculate the trace purity of the emitted single-photon state
\begin{equation}
\mathbb{P}=\iint\mathop{\text{d}t_1}\mathop{\text{d}t_1'} \left|\braket{t_1'|\chi(t\rightarrow\infty)|t_1}\right|^2/{P_1^2}.
\end{equation}
The limit $t\rightarrow\infty$ ensures the photons are completely emitted, and the single-photon elements of the density matrix are computed according to Eq. \ref{eq:chi_tpt}, i.e.
\begin{eqnarray}
&&\braket{t_1'|\chi(t\rightarrow\infty)|t_1} =\\
&&\quad\text{tr}_\text{sys}\big[\mathcal{K}(\infty,\tilde{\tau}_2)\mathcal{S}_{Q[\tilde{\tau}_{2}]}\mathcal{K}(\tilde{\tau}_{2},\tilde{\tau}_{1})\mathcal{S}_{Q[\tilde{\tau}_{1}]}\mathcal{K}(\tilde{\tau}_1,0)\ket{\psi}\bra{\psi}\big].\nonumber
\end{eqnarray}
Meanwhile, the single-photon emission probability is computed from a special case of Eq. \ref{eq:Pn__0} with $n=1$. The results for $\mathbb{P}$ are shown as the green curve in Fig. \ref{fig:4}, and unlike the visibility (blue curve), the trace purity of the single-photon state is unity for zero dephasing as expected. Further, as the dephasing rate $\gamma_1$ increases, it can be seen that the single-photon state becomes increasingly more mixed due to the drop in trace purity. Although the visibility follows the same basic trend as the trace purity they can be seen to have different dependencies and have significant disagreement.

From an experimental perspective, the significance of this discussion is to understand that the Hong--Ou--Mandel visibility is not identical to the trace purity (except for an ideal single-photon source) and that the visibility parameter actually is dependent on photon re-excitation just like the Hanbury--Brown and Twiss experiment. From a theoretical perspective, we hope that by providing a way to directly access the photonic density matrix, other metrics for quantifying single-photon source indistinguishability can become practical. For example, it has been suggested that the Frobenius distance metric $||\chi_\text{source}-\chi_\text{target}||^2$ might provide a better definition for indistinguishability \cite{walmsley2007generation}.

\section{Conclusions}

In summary, we have provided a complete framework for understanding zero-dimensional Hamiltonians as emitters of bosonic particles such as photons or phonons. Of practical relevance is that our formulation allows for the inclusion of dissipation into the particle emitters' dynamics. Because dissipation is often present in physical sources of particles it is important to model correctly for applications in quantum information processing. Although we only considered static dephasing, we believe our techniques are already applicable to power-dependent dephasing \cite{mccutcheon2010quantum} and that it might be possible to extend them to a polaron theory \cite{mccutcheon2010quantum,manson2016polaron,gustin2017influence} as well. Finally, our formalism ties together nearly all aspects of Markovian open-quantum systems, and reveals the connections between $(0+1)$-$d$ field theories, continuous matrix product states, and quantum stochastic calculus.

\section{Acknowledgements}

We thank Jelena Vu\v{c}kovi\'c and Shuo Sun for feedback and discussions, and we gratefully acknowledge financial support from the National Science Foundation (Division of Materials Research Grant No. 1503759) and the Air Force Office of Scientific Research (AFOSR) MURI Center for Quantum Metaphotonics and Metamaterials (Award No. FA9550-12-1-0488) and MURI Center for Attojoule Nano-Optoelectronics (Award No. FA9550-17-1-0002). D.L. acknowledges support from the Fong Stanford Graduate Fellowship and the National Defense Science and Engineering Graduate Fellowship. R.T. acknowledges support from the Kailath Stanford Graduate Fellowship.

\appendix

\section{Particle emission into fermionic reservoirs}

In the main text, we considered the reservoirs to have a bosonic character, meaning that the continuous field mode operators obeyed the commutation relations $[b(t),b^\dagger(s)]=\delta(t-s)$. Because the interaction Hamiltonian is linear in field operators, however, only one particle can be emitted in any time interval $\text{d}t$. Hence, the bosonic character of the reservoirs is actually never used in the problem. To emphasize this point, further consider the case where the reservoirs have a fermionic character so the continuous-mode operators anti-commute $\{b(t),b^\dagger(s)\}=\delta(t-s)$, as was done in \citet{haack2015continuous}. The key step that changes is from Eq. \ref{eq:commutestuff}: the commutator becomes an anti-commutator with the result that $\{\mathds{1}\otimes b(t_1),U(s,0)\}=U(s,t_1)\left(L\otimes\mathds{1}\right)U(t_1,0)$ for $s>t_1$. Applying this relation to the expansion of $U(t)$ and using the anti-commutation relations of $b(t)$ yields identical results through the rest of the paper. As expected, in order to distinguish between emission into bosonic or fermionic reservoirs, the system-reservoir coupling must be non-linear in field operators.

\section{List of important superoperators}
\begin{itemize}
\item $\mathcal{U}(t,0)\rho\equiv U(t,0) \rho\, U(0,t)$ is the unitary evolution superoperator.
\item $\mathcal{S}_0\chi=L_0\chi$ and $\mathcal{S}_1\chi=\chi L^\dagger_0$ are simply used to pre- or post-multiply the coupling operator $L_0$. $\mathcal{J}[L]\rho=L\rho L^\dagger=\mathcal{S}_0\mathcal{S}_1\rho$ is the recycling or emission superoperator.
\item $\rho(t_1)=\mathcal{M}(t_1,t_0)\rho(t_0)$ represents Lindblad evolution of the system density matrix, generated by the Liovillian $\mathcal{L}$.
\item $\mathcal{V}(t,0)\chi\equiv V(t,0) \chi V(0,t)$ with $V(t_1,t_0) =\bra{\bm{0}}U(t_1,t_0)\ket{\bm{0}}_\text{wg}$ represents evolution conditioned on no emission into the waveguide.
\item $\mathcal{K}(t_1,t_0)=\text{tr}_\text{bath}\big[\mathcal{V}(t_1,t_0)\{\ket{\psi}\bra{\psi}\otimes\ket{\bm{0}}\bra{\bm{0}}_\text{bath}\}\big]\nonumber$ is an unnormalized map that evolves the density matrix conditional on no photon emission into the waveguide.
\end{itemize}


\bibliography{bibliography}

\end{document}